\documentclass[twocolumn,aps,amsmath,amssymb,color,superscriptaddress,prl,footinbib,longbibliography]{revtex4-2}
\usepackage{stix}
\usepackage[colorlinks,citecolor=blue,linkcolor=blue,urlcolor=blue]{hyperref}
\usepackage{graphicx}
\usepackage{dcolumn}
\usepackage{multirow}
\usepackage{bm}
\usepackage{times}
\usepackage{tabularx}
\usepackage{footmisc}
\usepackage{xcolor}
\usepackage{float}
\usepackage{tabularx}
\usepackage{url}
\usepackage{hyperref}
\usepackage{ulem}
\usepackage{titlesec}
\usepackage[english]{babel}

\newcommand{\mb}{\mathbf}

\begin{document}

\title{Spontaneous Symmetry Breaking of Cavity Vacuum and Emergent Gyrotropic Effects in Embedded moir\'{e} Superlattices}

\author{Zuzhang Lin}
\affiliation{New Cornerstone Science Lab, Department of Physics, The University of Hong Kong, Hong Kong, China}
\affiliation{HK Institute of Quantum Science \& Technology, University of Hong Kong, Hong Kong, China}

\author{Hsun-Chi Chan}
\affiliation{New Cornerstone Science Lab, Department of Physics, The University of Hong Kong, Hong Kong, China}
\affiliation{HK Institute of Quantum Science \& Technology, University of Hong Kong, Hong Kong, China}

\author{Wenqi Yang}
\affiliation{New Cornerstone Science Lab, Department of Physics, The University of Hong Kong, Hong Kong, China}
\affiliation{HK Institute of Quantum Science \& Technology, University of Hong Kong, Hong Kong, China}

\author{Yixin Sha}
\affiliation{HK Institute of Quantum Science \& Technology, University of Hong Kong, Hong Kong, China}

\author{Cong Xiao}
\affiliation{Institute of Applied Physics and Materials Engineering, University of Macau, Taipa, Macau, China}

\author{Shuang Zhang}
\affiliation{New Cornerstone Science Lab, Department of Physics, The University of Hong Kong, Hong Kong, China}
\affiliation{HK Institute of Quantum Science \& Technology, University of Hong Kong, Hong Kong, China}

\author{Wang Yao}
\email{wangyao@hku.hk}
\affiliation{New Cornerstone Science Lab, Department of Physics, The University of Hong Kong, Hong Kong, China}
\affiliation{HK Institute of Quantum Science \& Technology, University of Hong Kong, Hong Kong, China}

\date{\today}
\begin{abstract}
In an electronic system, spontaneous symmetry breaking can arise from many-body interaction between electrons, leading to degenerate ground states distinguishable by emergent effects otherwise prohibited by the symmetry. Here we show that ultrastrong coupling of a mesoscopic electronic system to the vacuum of a cavity resonator can lead to another paradigm of spontaneous breaking of spatial symmetries in both systems. As a pertinent example, we consider the orbital gyrotropic effects in a moir\'{e} superlattice embedded in a THz split ring cavity resonator. Our mean-field and exact diagonalization calculations consistently demonstrate a spontaneous parity symmetry breaking in both the electronic ground state and the cavity vacuum, leading to two degenerate hybrid ground states distinguished by their opposite orbital gyrotropic Hall and magnetic effects.
These sizable responses in the cavity-embedded moir\'{e} superlattice are highly tunable by both the cavity field polarization and interlayer bias on the moir\'{e} superlattice, providing an advanced platform for manipulating gyrotropic effects.
\end{abstract}
\maketitle

Electronic systems respond to external electric and magnetic fields in ways that are restricted by their inherent symmetries~\cite{dresselhaus_group_2008,cohen2016fundamentals}. The responses can be controlled by breaking the symmetries through applying strain, introducing defects, creating materials interface, etc., which in general lead to a definite form of the response of interest. Coulomb interaction between electrons can also spontaneously break symmetries by establishing many-body correlations at low temperature, leading to a degenerate set of ground states each with its unique responses that are originally prohibited.

Cavity vacuum control has recently emerged as a new frontier in the manipulation of electronic properties, exploiting the vacuum quantum fluctuations which can dress electronic states with virtual cavity photons. Experiments observed significant influences of vacuum fluctuations on a broad range of phenomena, such as integer and fractional quantum Hall effects \cite{appugliese_breakdown_2022, enkner2024enhanced}, metal-to-insulator transition~\cite{jarc_cavity-mediated_2023}, magneto-transport~\cite{paravicini-bagliani_magneto-transport_2019}, and chemical reaction \cite{thomas_tilting_2019}. The exchange of virtual photons in the cavity vacuum can also mediate a plethora of electron correlation phenomena \cite{frisk2019ultrastrong},
including superconductivity \cite{schlawin2019cavity,gao2020photoinduced}, superfluidity \cite{schlawin2019pairin}, and charge-density-wave \cite{li2020manipulating}.
Theoretical studies have also shown that vacuum fluctuation is capable of transmitting broken symmetries of cavity confinements to the electronic system~\cite{sedov2022cavity,ke2023vacuum,vinas2023controlling}.
The intriguing prospect of spontaneous symmetry breaking in both the cavity field and electronic systems, which could enable otherwise prohibited responses, remains an unexplored area.

Moir\'{e} superlattices of 2D atomic crystals embedded in metallic split-ring terahertz (THz) electromagnetic resonator can provide diverse opportunities for the exploration of cavity vacuum control~\cite{lin2023remote}. The deep sub-wavelength confinement in the micron scale ring gap leads to dramatic enhancement of the vacuum quantum fluctuations in the ultra-strong light-matter coupling regime~\cite{appugliese_breakdown_2022,ciuti2021cavity,arwas2022quantum,scalari_ultrastrong_2012,chen_review_2016,keller_few-electron_2017,jeannin2019ultrastrong,paravicini-bagliani_gate_2017,paravicini-bagliani_magneto-transport_2019}.
The formation of superlattices through twisting controlled moir\'{e} patterns leads to versatile engaging material properties and meV-scale bandwidth and bandgap, ideal for exploring the various aspects of vacuum control by THz cavities.
The cavity mode is well discretized in energy while spanning all over superlattice sites in its micron scale volume, defining a mesoscopic system where many electrons are coupled to a common fundamental cavity mode. This mesoscopic nature underlies the nonlocal character of the virtual photon mediated electron-electron interaction, leading to intriguing possibilities such as remote control of topological matter~\cite{lin2023remote}.

In this work, we present a paradigmatic example of spontaneous symmetry breaking simultaneously in the vacuum of a split-ring cavity resonator and in the ground state of an embedded mesoscopic moir\'{e} superlattice.
We consider orbital gyrotropic Hall effect (OGHE) from Berry curvature dipole and orbital gyrotropic magnetic effect (OGME) from orbital magnetic moment dipole, which require breaking both the $C_3$ rotational symmetry and parity symmetry of the moir\'{e} superlattice.
Employing mean-field calculation and exact diagonalization (ED), we find the ground states of the hybrid system display spontaneous symmetry-breaking, with a two-fold degeneracy marked by distinct gyrotropic tensors and cavity field, related by parity transformation.
Notably, the vacuum field-induced gyrotropic effects are highly tunable by the polarization of the cavity field and the interlayer bias applied to the moir\'{e} superlattice. We emphasize that the emergence of the gyrotropic effect here is a novel symmetry breaking phenomenon arising from strong vacuum fluctuation, which imprints the $C_3$ symmetry breaking of cavity confinement to the moir\'{e} and triggers the spontaneous parity symmetry breaking in both the cavity field and the moir\'{e}.
The dual role of cavity vacuum fluctuation can be generally explored in other symmetry-breaking physics.

\begin{figure}
\centering
\includegraphics[width=1\columnwidth]{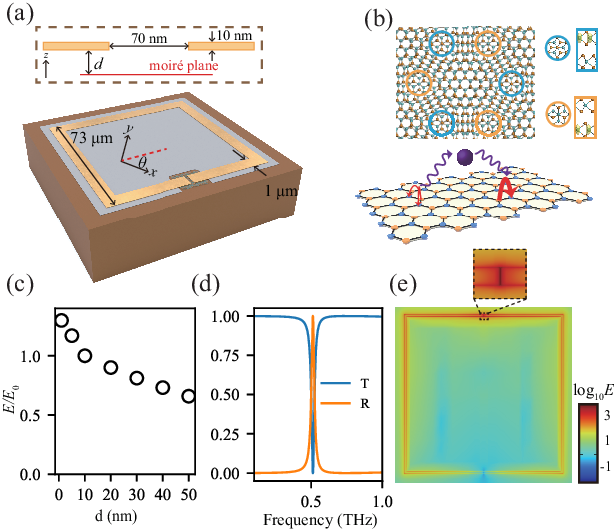}
\caption{ (a) Setup of the cavity-embedded moir\'{e} superlattice (the lower panel) encompassing a planar metallic split-ring Thz resonator (the gold ring) deposited on top of an h-BN substrate (the dark gray sheet). The upper panel shows a lateral perspective of the setup. 
(b) Illustration of the hexagonal moir\'{e} superlattice (the upper panel) composed of two moiré orbitals localized in the top (cyan circles) and bottom layers (orange circles), and the schematic diagram of the $C_3$ rotational symmetry breaking (the lower panel) as a result of distinct cavity vacuum field effects on holes' hopping across various directions. The thickness of the red arrows signifies the magnitude of the impact of the cavity vacuum field. 
(c) The electric field $E$ averaged over a rectangular area of 0.3 \textmu m $\times$ 1 \textmu m, as a function of the vertical distance $d$. $E_0$ represents the value at $d=10$ nm.
(d) Resonator's reflection and transmission spectra. (e) The cavity's electric field (in V/m) profile on the plane at a vertical distance $d= 20$ nm from the split-ring. The inset displays an enlarged view close to the gap.
}
\label{Fig-Conf}
\end{figure}

\textit{Vacuum field-induced gyrotropic effects.}
 Gyrotropic effects generally describe the linear response of a time-reversal-odd axial vector with respect to a time-reversal-even polar vector in nonmagnetic crystals. The response coefficient $\alpha_{ab}^{i}$ (we denote $i=$ H for OGHE and M for OGME), the so called gyrotropic tensors, can only be time-reversal even in nonmagnetic systems \cite{Culcer2007}. As such, the effects must involve carrier relaxation processes on the Fermi surface, and $\alpha^{i}_{ab}$ starts from the first order in relaxation time $\tau$: $\alpha_{ab}^i = \tau R^i_{ab}$. Here $R^i_{ab}$ is intrinsic to the band structures and reads \cite{sodemann_quantum_2015,zhong_gyrotropic_2016,Pesin2019,lu2021}
\begin{equation}\label{Eq-formula}
R^i_{a b}=\int_{\mathrm{BZ}} \frac{d^2 k}{(2 \pi)^2} v_a \frac{d f}{d \varepsilon_k} B^i_{\mathbf{k}, b},
\end{equation}
where $a, b=x, y, z$, $\mathbf{v}$ is the band velocity, $f$ is the Fermi-Dirac distribution function, and $\varepsilon_{\boldsymbol{k}}$ is the band energy. For OGHE (OGME ),
$R^{\mathrm{H}}$ ($R^{\mathrm{M}}$) represents the Berry curvature dipole (orbital magnetic moment dipole), with $B^{\mathrm{H}}$ ($B^{\mathrm{M}}$) symbolizing the momentum space Berry curvature  (orbital magnetic moment).
In two-dimension, $a=x,y$ ($a$ represents the direction of the external field) and $b=z$ ($z$ is set to perpendicular to the moir\'{e} plane) in Eq. (\ref{Eq-formula}) and $R_{a b}^i$ can be abbreviated as $R_a^i$ for brevity.
Obviously, $R_a^i$ is forbidden under parity and any rotational symmetry along $z$ axis.

The moir\'{e} superlattice by itself preserves the $C_{3z}$ rotational symmetry, thus prohibiting any tensor elements $R^i_{a}$. When placed in the proximity of the cavity gap, carrier hopping in the superlattice is influenced by the cavity vacuum field (Fig. \ref{Fig-Conf}), which breaks the $C_{3z}$ symmetry.
Nevertheless, both the moir\'{e} and the cavity mode here have the parity symmetry, which still forbids $R^i_{a}$.
The gyrotropic
effects can only emerge from spontaneous parity symmetry breaking, as the collective interplay of many carriers in the mesoscopic moir\'{e} superlattice and the virtual cavity photon, as demonstrated by the mean-field and exact diagonalization results in the following sections.

\begin{figure}
\centering
\includegraphics[width=1\columnwidth]{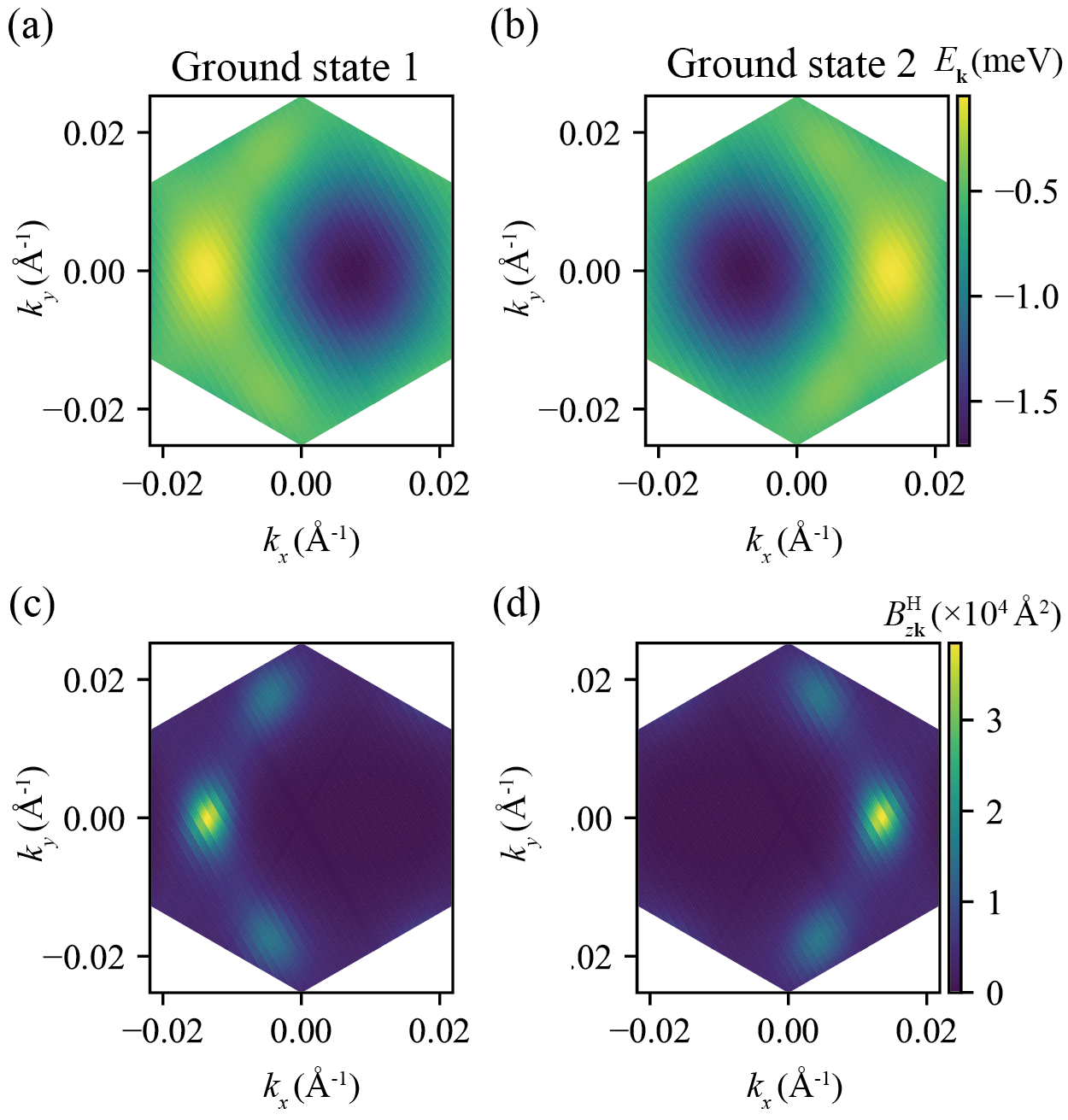}
\caption{(a), (b) Mean-field dispersions for the lower miniband of the cavity renormalized moir\'{e} superlattice in its two degenerate ground states respectively. (c), (d) The corresponding band Berry curvature ($B^{\rm H}_{z\mathbf{k}}$) in (a) and (b).}
\label{Fig-Energy}
\end{figure}

\textit{Model of the cavity-embedded moir\'{e} superlattice.}
We consider a single-gap resonator
with a geometry schematically shown in Fig.~\ref{Fig-Conf}(a)].
The resonant frequency for the fundamental cavity mode is 0.512 Thz [Fig.~\ref{Fig-Conf}(d)]. The mode has an electric field distribution highly concentrated in the vicinity of the gap region [c.f. Fig.~\ref{Fig-Conf}(e)], where
the moir\'{e} superlattice can be positioned to experience a strong vacuum field.
The simulation of the THz resonator is performed via COMSOL Multiphysics software and the details can be found in the Section VI of the Supplemental Material.

The moir\'{e} superlattice under consideration is the homobilayer transition metal dichalcogenides with a small twist from the R-type stacking [Fig.~\ref{Fig-Conf} (b)]. Its lowest energy valence states at the $\mb{K}$ ($\mb{K}^{\prime}$) valley  can be effectively represented by a two-band tight-binding (TB) model on a honeycomb superlattice, with complex amplitude for the next-nearest-neighbor hopping originating from the real-space Berry curvature~\cite{yu2020giant}, which is essentially the Haldane model. We consider only one valley, as the other valley contributes equivalently to the gyrotropic effects \cite{example}.
The Hamiltonian of the  superlattice coupling to the cavity field reads,
 \begin{equation}\label{eq_tb}
\begin{aligned}
H_{\mathrm{m}} & =\sum_{\boldsymbol{k} \mb{\delta}}\left(t_1 e^{i \boldsymbol{k} \cdot \boldsymbol{\delta}} e^{i \chi \xi_\delta\left(a+a^{\dagger}\right)} c_{B \boldsymbol{k}}^{\dagger} c_{A \boldsymbol{k}}+\text { h.c. }\right) \\
& +\sum_{k \boldsymbol{\tau} \mu}\left(t_2 e^{i \phi_\mu} e^{i \boldsymbol{k} \cdot \boldsymbol{\tau}} e^{i \chi \xi_\tau\left(a+a^{\dagger}\right)} c_{\mu \boldsymbol{k}}^{\dagger} c_{\mu \boldsymbol{k}}+\text { h.c. }\right)
\end{aligned}
\end{equation}
where
$\mu=A,B$ denote the two sublattice sites, $t_1$ and $t_2$ respectively refer to the nearest- and next-nearest-neighbor hopping strength, $\phi_{\mu}$ is the phase of the next-nearest-neighbor hoping, 
$\boldsymbol{\delta}$ and $\boldsymbol{\tau}$ are respectively the nearest- and next-nearest-neighbor hoping vectors. 
$\phi_{\mu}=2\pi/3$ for $\mu=A$ and $-2\pi/3$ for $\mu=B$.
The cavity field vector potential is $\mb{A}=A_0\boldsymbol{e}_p(a+a^{\dagger})$, with $a^{\dagger}$ ($a$) the photon creation (annihilation) operator, and $A_0$ and $\mb{e}_p$ the amplitude and directional unit vector of the cavity field, respectively.
The coupling strength $\chi$ is defined as $eA_0|\boldsymbol{\delta}|/\hbar$, and
$\xi_{\boldsymbol{\delta}}=\boldsymbol{e}_p \cdot \frac{\boldsymbol{\delta}}{|\boldsymbol{\delta}|}$, $\xi_{\boldsymbol{\tau}}=\boldsymbol{e}_p \cdot \frac{\boldsymbol{\tau}}{|\boldsymbol{\delta}|}$.
The Hamiltonian $H_{\mathrm{m}}$ is invariant under the parity transformation.

To capture the low-energy physics of the Hamiltonian in Eq. (\ref{eq_tb}), we  employ  the Schrieffer-Wolff transformation \cite{schrieffer_relation_1966} to  project the total Hamiltonian $H_{\text{tot}}=H_{\text{m}}+H_{\text{c}}$ ($H_{\text{c}}=\hbar \omega a^{\dagger} a$ is the bare photon Hamiltonian) to the lowest-energy subspace (photon number $n=0$). The effective Hamiltonian reads
 \begin{equation}\label{eq-mf}
H_0^{\mathrm{eff}}=\langle 0|H_{\mathrm{m}}| 0 \rangle+\sum_{l>0} U_l V_l,
\end{equation}
where $U_l=\sum_{\mathbf{k} \mu \nu} P_{\mathbf{k} \mu}\langle 0|H_{\mathrm{m}}|l\rangle P_{\mathbf{k}\nu} l \hbar \omega /[(E_{\mathbf{k} \mu}-E_{\mathbf{k} v})^2-(l \hbar \omega)^2]$, $V_l=\sum_{\mathbf{k} \mu \nu} P_{\mathbf{k} \mu}\langle l|H_{\mathrm{m}}| 0\rangle P_{\mathbf{k} \nu}$ with $P_{\mathbf{k} \mu}=|\mathbf{k} \mu \rangle\langle\mathbf{k} \mu|$ the projection operator, $l$ is the photon number index and $E_{\mathbf{k} \mu, l}=E_{\mathbf{k} \mu}+l \hbar \omega$. More details can be found in Sections II and III of the Supplemental Material.

\textit{Results from mean-field and exact diagonalization calculations.}
The term of product of four fermionic operators ($U_l V_l$) represents the interaction between holes via the exchange of virtual cavity photons.
These terms can be treated using mean-field approximation, which leads to a mean-field Hamiltonian
 $H_{\mathrm{mf}} \approx\langle 0|H_{\mathrm{m}}| 0\rangle+\sum_{l>0}(\left\langle U_l\rangle V_l+U_l\langle V_l\rangle-\langle U_l\rangle\langle V_l\rangle\right)$  that must be solved self-consistently (details can be found in Section III of the Supplemental Material).
 It is noted that  the Hamiltonian of Eq. (\ref{eq-mf}) is invariant under parity transformation, adhering to the system's parity symmetry. Yet, the mean-field Hamiltonian experiences a spontaneous parity symmetry  breaking when the nonzero order parameters  $\langle U_l\rangle$ and $\langle V_l\rangle$ are solved self-consistently.

Based on the established mean-field Hamiltonian, we next perform calculations to demonstrate the vacuum field-induced gyrotropic effects. We consider the moir\'{e} superlattice of a mesoscopic size in a rectangular region with dimensions of 0.3 \textmu m by 1 \textmu m. Assuming the lattice plane has a vertical distance of $d=20$ nm from that of the metal ring, the amplitude of the cavity electric field at the lattice can lead to a $\chi \approx 0.03$ (see more details in the Section VI of Supplemental Material).
In the mean field calculations,
the hopping amplitudes are taken as $t_1=0.35$ meV and $t_2=0.1$ meV, the derivative of Fermi Dirac distribution $\frac{df}{d\epsilon}$ is taken as a Gaussion function of width 0.15 meV, and we consider a filling factor of 0.9 hole per supercell.

\begin{figure}
\centering
\includegraphics[width=1\columnwidth]{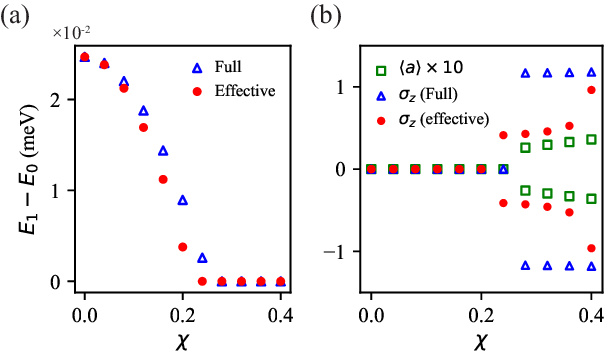}
\caption{Exact diagonalization calculations. (a) Difference between the two lowest energy eigenvalues, and (b) orbital polarization (c.f. main text), with the increase of cavity coupling strength $\chi$.  Solid red circles are from ED calculations of the effective Hamiltonian of cavity-mediated hole-hole interaction (Eq. (\ref{eq-mf})).
Hollow blue triangles
are from ED of the full Hamiltonian of superlattice-cavity coupled system (Eq.~(\ref{eq_tb})). ED of the full Hamiltonian also gives the expectation value of cavity field $\langle a \rangle$, shown by hollow green squares in (b).
}
\label{Fig-ED}
\end{figure}

Remarkably, the mean-field ground state features a two fold degeneracy.
The energy dispersion of the mean-field bands differ a lot for the two ground states [Fig. \ref{Fig-Energy}(a) and (b)] and obviously break $C_3$ rotational symmetry.
The distinct band dispersions for the two ground states are further accompanied by their different momentum-space Berry curvature ([Fig. \ref{Fig-Energy}(c) and (d)]) and orbital magnetic moment (Fig. S4 of the Supplemental Material). Notably, Berry curvature dipoles and orbital magnetic moment dipole are not only nonzero but opposite for these two ground states, which suggests that the parity symmetry is spontaneously broken in the ground states, and the two degenerate ground states are related by the parity transformation
(more details in the Section IV of the Supplemental Material).

To further validate the proposition that a spontaneous parity symmetry breaking can indeed develop from the parity-symmetric cavity-mediated hole-hole interaction Eq.~(\ref{eq-mf}) or the full Hamiltonian of superlattice-cavity coupled system Eq.~(\ref{eq_tb}), we also perform exact diagonalization (ED) calculations with both Hamiltonians on a 3 by 5 unit-cell cluster on a torus geometry. The hole number is set to 14. Upon increasing the coupling strength $\chi$, the difference between the two lowest energy eigenvalues gradually decreases [Fig. \ref{Fig-ED}(a)] and vanishes for $\chi > 0.24$, indicating a two-fold degeneracy of the many-body ground state. The observed double degeneracy in ED calculations validates the one identified in mean-field calculations.   Meanwhile, the orbital polarization $\sigma_z=\sum_{\mathbf{k}}\langle c^{\dagger}_{A\mathbf{k}}c_{A\mathbf{k}}-c^{\dagger}_{B\mathbf{k}}c_{B\mathbf{k}}\rangle$ becomes nonzero and is contrast for the two degenerate states [Fig. \ref{Fig-ED}(b)]. Since orbital polarization is prohibited by parity symmetry, it can serve as an indicator for parity symmetry breaking.
ED calculations based on the
cavity-mediated hole-hole interaction Hamiltonian Eq.~(\ref{eq-mf}) and that based on the full Hamiltonian of superlattice-cavity coupled system Eq.~(\ref{eq_tb}) agree well.
ED of the full Hamiltonian also gives opposite expectation value $\langle a \rangle$ in the two degenerate ground states, showing the parity symmetry breaking in the cavity vacuum field.
A more detailed discussion of ED calculations can be found in Section V in the supplementary material.

\begin{figure}
\centering
\includegraphics[width=1\columnwidth]{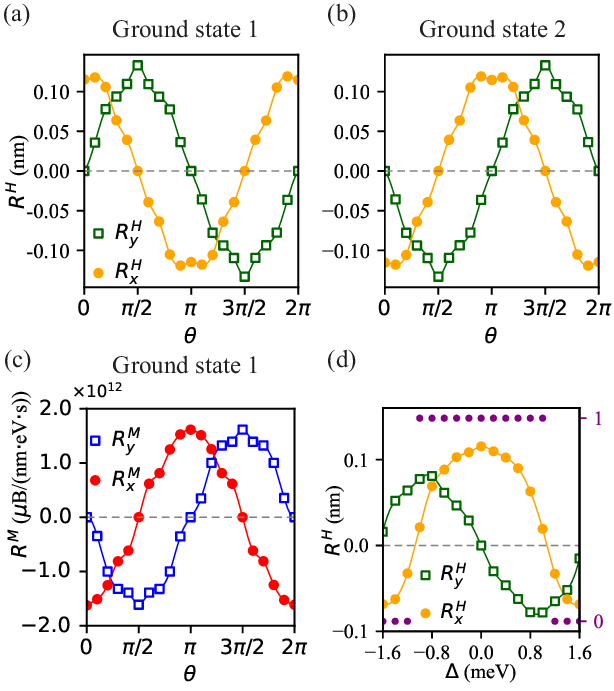}
\caption{(a)-(b) Dependence of Berry curvature dipole on cavity field polarization for two degenerate ground states, when the interlayer bias applied to the moir\'{e} is set to zero. (c) The GME tensor $\Gamma$ vs cavity field polarization for ground state 1, using the same calculations parameters of (a) and (b). (d) Dependence of Berry curvature dipole on interlayer bias when cavity field polarization is set to along the $x$ axis ($\theta=0$). The purple scatters refer to the Chern number.  In all calculations, the filling factor $f$ is maintained at 0.9, and scatter points are calculated, while the connecting lines merely enhance readability.
}
\label{Fig-bias}
\end{figure}

Upon determining the gyrotropic effects for a given set of parameters, we proceed to investigate the dependence of the $R_a$ on the direction of the cavity field polarization and the onsite energy difference. Denote the cavity polarization as $e_p=(\cos \theta, \sin \theta)$ with $\theta=0$ corresponding to the polarization pointing along the $x$ direction. The gyrotropic tensor $R_a$ exhibits a sinusoidal-like dependence on the variable $\theta$ [Figs. \ref{Fig-bias}(a)-\ref{Fig-bias}(c)], resulting in both $R_x$ and $R_y$ encompassing positive and negative values. Throughout the range of the variable $\theta$, an inverse relationship between $R_a$ for two ground states is consistently observed.
Intriguingly, it turns out that when the cavity field polarization is adjusted to along the $y$ direction ($\theta=\pi/2+N\pi$ with $N$ as an integer), we find $R_x=0$ but $R_y \neq 0$, whereas for cavity polarization along the $x$ direction ($\theta=N\pi$),  $R_y=0$ but $R_x \neq 0$. The underlying mechanism is that the system upholds $C_{2x} $ and $C_{2y} $ for the former and latter case, and thus forbids $R_x$ and $R_y$, respectively. Therefore, the occurrence of solely one gyrotropic tensor component (either $R_x$ or $R_y$) for a particular cavity configuration is realized.

The emergent gyrotropic effects can be further controlled by a modest interlayer bias $\Delta$, which introduces on-site energy difference between the two sublattices of the moire polarized in opposite layers [c.f. Fig. 1(b)]. Figure~\ref{Fig-bias}(d) illustrates the interlayer bias dependence of the Berry curvature dipole.
Clearly, when a finite $\Delta$ is introduced, $R^{H}_y$ emerges due to the breaking of the $C_{2y}$ symmetry, and the amplitude of $R^{H}_x$  simultaneously changes. One notices that the ground state degeneracy is absent at finite interlayer bias, which explicitly break the parity symmetry.
All gyrotropic tensor components in the ground state under a given interlayer bias is inversely related to those under the opposite interlayer bias.
Moreover, $\Delta$ can induce a topological phase transition in the moire superlattice. Remarkably, we find the topological transition is always accompanied by a continuous sign change of the $R^{H}_x$, as shown by the example in Fig.~4(d).

\textit{Discussion and conclusion.}
We note that in the literature exploring superradiant phase transition for the cavity vacuum, the possible occurrence of photon condensation marked by nonzero $\langle a \rangle$ has been extensively debated \cite{hepp1973superradiant,wang1973phase,nataf_no-go_2010,rzazewski1975phase, viehmann_superradiant_2011-1,andolina_cavity_2019}.
In contrast to such phase that is characterized by the presence of ground-state coherent photons in a macroscopic system, our focus inherently deviates towards an entirely different scenario, namely, a mesoscopic configuration where the thermodynamic limit is not applicable. Our ED calculation on a mesoscopic system unambiguously demonstrates the occurrence of nonzero $\langle a \rangle$, as a signature of the spontaneous parity symmetry breaking in the cavity vacuum.



In conclusion, we have shown that strong light matter interaction in a moir\'{e} superlattice embedded in the vacuum of a split-ring THz resonator can lead to a spontaneous parity symmetry-breaking in the cavity vacuum and emergent gyrotropic effects of the embedded moir\'{e} superlattice.  The vacuum field-induced gyrotropic effects possess significant magnitudes as compared to existing mechanisms in 2D materials~\cite{ma_observation_2019,kang_nonlinear_2019,son_strain_2019,lesne_designing_2023}, and can be effectively modulated by cavity field polarization and the interlayer bias, underlying a remarkable tunability as compared to those mechanisms relying on strain or introducing interface to break symmetry.
Furthermore, spin-orbit coupling is absent in this vacuum field-induced gyrotropic effects, despite the prevailing notion that gyrotropic effects generally necessitate the spin orbit coupling.

\textbf{Acknowledgment:} The work is supported by Research Grant Council of Hong Kong SAR (A-HKU705/21, HKU SRFS2122-7S05, AoE/P-701/20), National Key R\&D Program
of China (2020YFA0309600), and New Cornerstone Science Foundation.


\begin{thebibliography}{41}
\bibitem [{exa()}]{example}%
  \BibitemOpen
  \href@noop {} {}\bibinfo {note} {Incorporating the other valley necessitates
  a TB model that encapsulates the physics of both valley, the primary effect
  of which is doubling the number of holes in comparison to the model detailed
  in the main text. The essence of the physics in this work can be addressed by
  integrating one valley.}\BibitemShut {Stop}%
\end{thebibliography}
%

\end{document}